\documentclass[conference]{IEEEtran}
\IEEEoverridecommandlockouts

\usepackage{cite}
\usepackage{amsmath,amssymb,amsfonts}
\usepackage{algorithmic}
\usepackage{hyperref}
\usepackage{graphicx}
\usepackage{textcomp}
\usepackage{xcolor}
\usepackage{comment}
\usepackage{multirow}  
\usepackage{booktabs}   

\def\BibTeX{{\rm B\kern-.05em{\sc i\kern-.025em b}\kern-.08em
    T\kern-.1667em\lower.7ex\hbox{E}\kern-.125emX}}
\begin{document}

\title{PGR-DRC: Pre-Global Routing DRC Violation Prediction Using Unsupervised Learning\\
\thanks{This work was supported in part by the National Science Foundation (NSF) award number: 2138253, and the UMBC Startup grant.}
}

\author{\IEEEauthorblockN{ Riadul~Islam and Dhandeep~Challagundla}
\IEEEauthorblockA{\textit{Department of Computer Science and Electrical Engineering} \\
\textit{University of Maryland, Baltimore County}\\
Baltimore, Maryland, USA \\
riaduli@umbc.edu, vd58139@umbc.edu}
}

\maketitle

\begin{abstract}
Leveraging artificial intelligence (AI)-driven electronic design and automation (EDA) tools, high-performance computing, and parallelized algorithms are essential for next-generation microprocessor innovation, ensuring continued progress in computing, AI, and semiconductor technology. Machine learning-based design rule checking (DRC) and lithography hotspot detection can improve first-pass silicon success. However, conventional ML and neural network (NN)-based models use supervised learning and require a large balanced dataset (in terms of positive and negative classes) and training time. This research addresses those key challenges by proposing the first-ever unsupervised DRC violation prediction methodology. The proposed model can be built using any unbalanced dataset using only one class and set a threshold for it, then fitting any new data querying if they are within the boundary of the model for classification. This research verified the proposed model by implementing different computational cores using CMOS 28 nm technology and Synopsys Design Compiler and IC Compiler II tools. Then, layouts were divided into virtual grids to collect about 60k data for analysis and verification. The proposed method has 99.95\% prediction test accuracy, while the existing support vector machine (SVM) and neural network (NN) models have 85.44\% and 98.74\% accuracy, respectively. In addition, the proposed methodology has about $26.3\times$ and up to $6003\times$ lower training times compared to SVM and NN-models, respectively.
\end{abstract}


\section{Introduction}

Early-stage Design Rule Checking (DRC) violation prediction is crucial in integrated circuit (IC) design as it significantly reduces engineering costs by identifying potential violations early in the design cycle. Traditional post-layout DRC checks can be computationally expensive and time-consuming, leading to multiple design iterations and increased turnaround time. By integrating predictive models or machine learning approaches, engineers can detect and resolve violations at the placement and routing (P\&R) stage, reducing the number of costly late-stage fixes and improving overall design efficiency~\cite{Liang_DRC_CNN:2022, islam2019soft, Mirhoseini_RL:2020, islam2022feasibility}. As long as the P\&R tools predict and resolve possible DRC violation hotspots before the global routing phase, it is possible to reduce significant engineering design time and cost~\cite{kahng2018no, esmaeilzadeh2024open}.

Moreover, early DRC violation prediction enables better resource allocation, reducing design tape-out risks and minimizing unnecessary fabrication costs. Recent advancements in artificial intelligence (AI)-driven design automation have demonstrated improved accuracy in predicting DRC hotspots, leading to faster design closure and optimized silicon performance~\cite{cnn_jhen:23, Lin_eDRC:2025, chen2025bi}.

Conventional AI-driven DRC violation prediction approaches either use machine learning (ML) or neural network (NN)- based models. ML-based models primarily collect features from different layers of physical design flow to build their models. On the other hand, NN-based models extensively use image-based features to train their models. However, all those conventional supervised models requires time-consuming training.

This research identified some key issues in conventional DRC violation prediction approaches; for example, if any design is divided into equal-sized grids, it is noticeable that the number of grids with DRC violations in the overall design is usually a lot less than the overall number of grids. In the case of supervised ML, there should be a reasonable balance between the positive (this research refers to as data with DRC violation) and negative (i.e., DRC violation-free data) samples. As a result, it is possible to model DRC violation prediction as an anomaly detection problem in which the training data does not focus on positive samples. Instead, it builds itself upon the anomaly-free data.

Besides, the feature characteristics of the DRC-violation-free grids exhibit similarities among different designs. It is shown in popular research work that unsupervised learning always provides an almost constant rate of prediction accuracy in terms of precision and recall, while the supervised learning model may fail in the case of a completely new scenario~\cite{Laskov:2005}. So, instead of including the DRC-violating data in the training model, this research fits a Gaussian distribution from only the DRC violation-free dataset. In that case, this research proposed an unsupervised methodology that can automatically identify an anomaly as a feature value from any new dataset. The key contributions of this research are,
\begin{itemize}
    \item First-ever unsupervised DRC violation prediction methodology.
    \item The proposed methodology requires $915\times$ lower training time compared to the existing convolutional neural network (CNN)-based DRC hotspot prediction method~\cite{cnn_jhen:23}.
    \item The proposed method has 14.51\% better accuracy than the existing ML-based model and 27.78\% better recall values compared to the existing CNN-based model~\cite{cnn_hung:23}.
     \item The proposed method has up to $26.3\times$ lower training time compared to conventional ML-based models~\cite{Chan_SVM:2016} and up to $6003\times$ lower compared to NN-based models~\cite{cnn_hung:23}.
\end{itemize}

The rest of the paper, organized as Section~\ref{sec:background} provides a brief introduction to existing ML-based physical design flow, Section~\ref{sec:proposed_method} will discuss the proposed methodology and step-by-step process of our new physical design flow, Section~\ref{sec:analysis} will provide detailed experimental setup, data analysis, and discussion on this research results compared with state of the art approaches, and finally Section~\ref{sec:conclusion} concludes the paper with final remarks and research contributions.
\section{Background}
\label{sec:background}


ML and NN-based techniques for predicting DRC violations require feature extraction at multiple stages of the P\&R flow. Some approaches utilize congestion reports generated after global routing~\cite{hung_transforming:20,cnn_hung:23,xie_routenet:18,yu_pin_deep_learning:19,almeida_hyperimage:25}, using machine learning and neural networks to predict DRC Violation (DRV) hotspots. However, since these techniques rely on global routing data, they do not contribute to reducing runtime overhead. Alternatively, other studies focus on extracting information post-placement~\cite{pgnn_park:24,cnn_jhen:23,Zhang_quantum:24,liang_jnet_drc:20,lin_deep_learning:24,Islam_cjece:2022,lu_yolov3:23,Islam_DAC:2019,Chan_SVM:2016, paul2023deep} to either directly predict DRVs or predict congestion maps of global routing to indirectly correlate with DRVs. Initial works primarily relied on information extracted from individual GCells, but subsequent advancements have expanded feature sets to include adjacent grid and macro information, significantly enhancing prediction accuracy and robustness.

Many studies employ traditional machine learning models for DRC prediction, including support vector machines (SVM)~\cite{Chan_SVM:2016,svm_chan:17,svm_chen:18}, random forest~\cite{Islam_DAC:2019, Islam_cjece:2022}, random under-sampling with Adaboost (RUSBoost)~\cite{tabrizi_rusboost:2017}, and multivariate adaptive regression splines (MARS)~\cite{chen_mars:2016,xie_routenet:18}. Other works explore more advanced neural network architectures such as multi-layer perceptrons (MLPs)~\cite{tabrizi_mlp:19}, convolutional neural networks (CNNs)~\cite{cnn_hung:23,cnn_jhen:23}, fully convolutional networks (FCNs)~\cite{lu_yolov3:23}, deep learning-based models like J-Net~\cite{liang_jnet_drc:20,lin_deep_learning:24}, or quantum-classical hybrid CNNs~\cite{Zhang_quantum:24} to enhance prediction accuracy and DRC hotspot detection.

The conventional CNN-based approach uses pre-global routing (PreGR)~\cite{cnn_jhen:23} or post-global routing (PostGR) image features~\cite{cnn_hung:23}  for training, as a consequence, requires a significant training time. 
Pin accessibility aware graph neural network (PA-GNN) combines GNN~\cite{Zhang_GNN:2020, Wang_GNN:2020} and U-net~\cite{Ronneberger_UNET:2015} to predict DRC violations. Unlike the PreGR CNN model, which uses existing ResNet~\cite{He_ResNet:2015} like architecture, FCN only uses convolutional layers to predict pixels' DRC violations. 

However, all those approaches use expensive supervised learning models.
This work uses a novel unsupervised learning model to predict DRC violations using features extracted from the pre-global routing stage of the P\&R design flow.

{
}


\section{Proposed Unsupervised DRC Violation Prediction Methodology}
\label{sec:proposed_method}
Unlike conventional electronic design and automation (EDA) approaches~\cite{islam2018low, Ding_clock:2023, Islam_rSRAM:2021, Islam_phd_thesis:2017}, whose primary objective is optimizing power~\cite{challagundla2024arxrcim, wang2024deep, challagundla2022power} and performance~\cite{Xing_timing:2023, islam2018dcmcs, Wang_timing:2023, guthaus2017current}, this research concentrates on ML-based physical design implementation acceleration. The first step of the proposed ML-based methodology is to build feature models extracted from the physical design flow. It is critical to characterize various feature distributions as Gaussian (normal).
The notation N ($\mu$, $\sigma ^2$) represents a Gaussian distribution with mean $\mu$ and variance $\sigma ^2$. This distribution is characterized by its bell-shaped curve, which is symmetric around $\mu$, indicating that data points are more likely to appear near the mean and less likely as they move further away. The standard deviation $\sigma ^2$ controls the spread of the distribution; a larger $\sigma ^2$ results in a wider curve, while a smaller $\sigma ^2$ leads to a narrower one. The probability density function (PDF) of the normal distribution is given by for a random variable $x$:
\begin{equation} 
\label{eq:pdf}
f(x; \mu, \sigma ^ 2) = \frac {1}{\sqrt {2\pi \sigma ^2}}e^(-\frac {(x-\mu)^2}{2 \sigma ^ 2})
\end{equation}
This distribution is fundamental in the Central Limit Theorem~\cite{casella2002statistical}, which states that the sum of many independent random variables tends to follow a normal distribution regardless of their original distributions. 

For each individual feature, we can compute the probability and multiply all the probabilities for $n$ features for each data sample, which is derived from the statistical independence assumption. Given a training set of samples, {$x^{(1)}$, $x^{(2)}$,\dots, $x^{(n)}$}, where each sample is a vector, we can compute the probability of a variable $x$ to be in the Gaussian distribution using the joint PDF factorizes as:
\begin{equation} 
\label{eq:pdf_coprod}
f(x) =  \prod_{i=1}^n  f(x_i; \mu _i, \sigma ^ 2_i) 
\end{equation} 
Once a threshold is determined for $f(x)$, this could be utilized to determine if the data point is within specific categories. If the value is within that threshold, then it will be considered a DRC violation-free datum. If $f(x) < threshold$, it will be predicted to be a DRC violation. An iterative method can be applied to identify the 
best threshold value corresponding to the best prediction accuracy.

\begin{figure}[t!]
\centerline{\includegraphics[width = 0.5\textwidth]{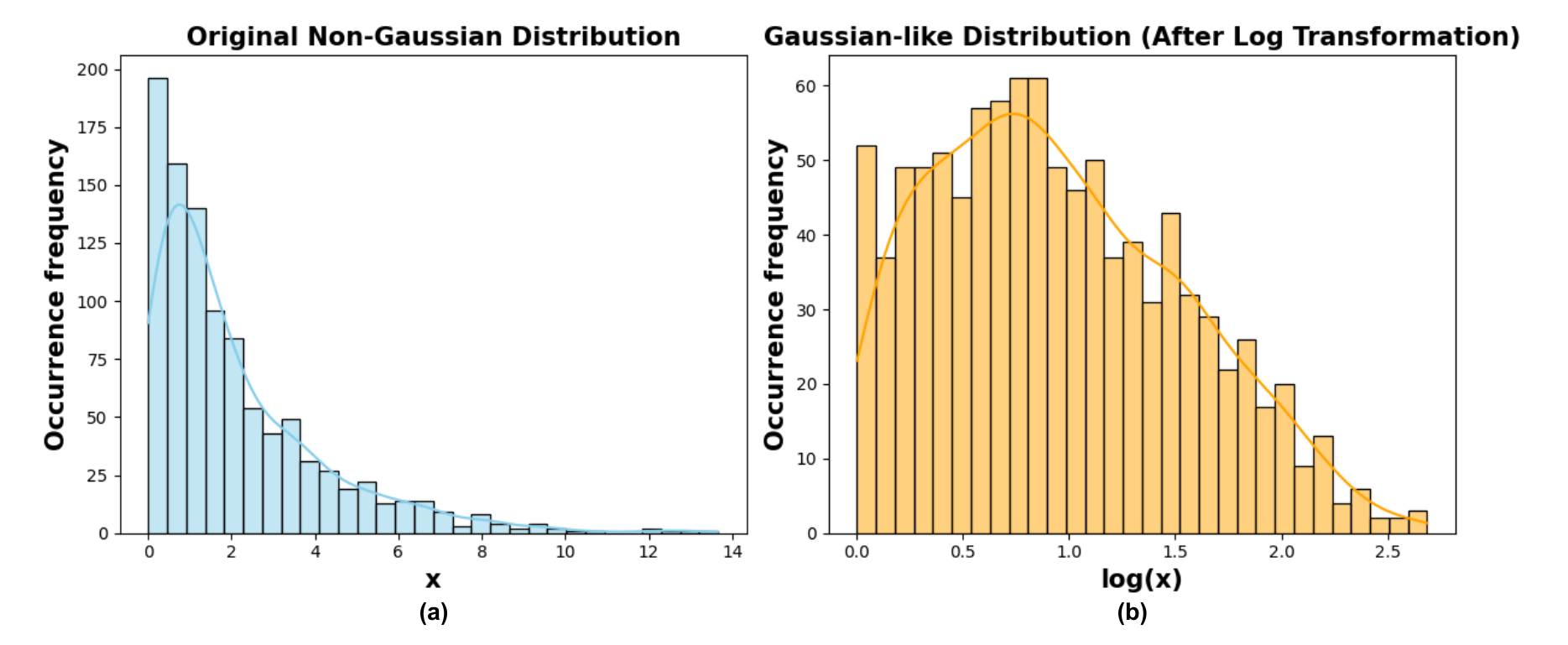}}
\vspace{-0.27cm}
\caption {Example of transforming a non-Gaussian distribution into a Gaussian distribution curve, (a) a non-Gaussian bar chart, and (b) a Gaussian distribution.}
\label{fig:gaussian_example}
\vspace{-0.05cm}
\end{figure}

{\bf Example of Gaussian Fitting:}
Fitting any feature in the proposed model requires a transformation of any given feature distribution into a Gaussian distribution. For example, we can plot the histogram of the feature values to check for the bell-shaped curve. If for any feature $x$, the distribution is not closely Gaussian, and it is possible to apply a standard mathematical transformation on that feature using log(x), log(x+1), log(x+constant), $\sqrt x$, etc. functions to transform it into a Gaussian distribution. Figure~\ref{fig:gaussian_example}(a) shows a non-Gaussian bar chart.
Now, applying a simple log transformation on this data, it is possible to obtain a bell-shaped curve as shown in Figure~\ref{fig:gaussian_example}(b), which is clearly a Gaussian distribution. In the proposed unsupervised DRC violation prediction framework, this research uses Python libraries,  
which have built-in functions to transform the features automatically into a Gaussian distribution.

\begin{figure}[h!]
\centerline{\includegraphics[width = 0.5\textwidth]{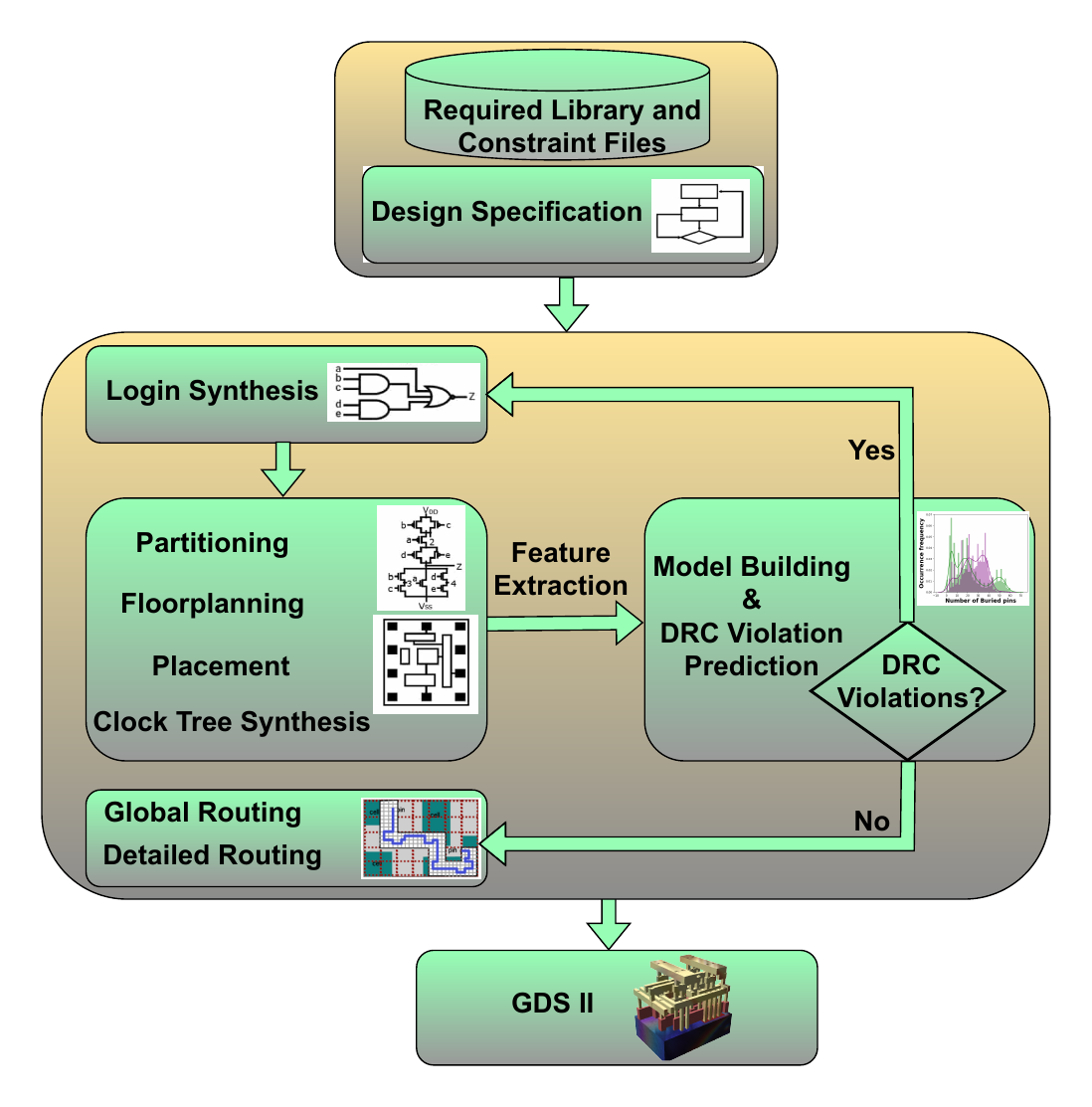}}
\vspace{-0.27cm}
\caption {The proposed methodology collects features before the detailed routing phase of the physical design flow and then builds an unsupervised model to predict DRC violations, improving the overall design implementation cost and efficiency of IC design.}
\label{fig:pd_flow}
\vspace{-0.05cm}
\end{figure}
{\bf Feature Selection:}
For the anomaly detection problem, there is a standard error analysis procedure. 
It is required that $f(x)$ is large for normal examples and small for anomalous examples. 
However, one common issue may arise when f(x) has an insignificant value for 
both types of examples. In this case, it is necessary to examine in detail the anomalous 
examples that are giving high probability and try to identify 
new features that will better distinguish the data. In other words, researchers 
have to choose features that show unusually different values in case 
of an anomaly. As a result, this research chooses empirically important features 
that make a significant contribution to DRC violation. 
 This research collected the following featured data from each grid after the layout was equally divided into grids for analysis to show the effectiveness of the proposed methodology: pin and cell densities, buried nets, buried cells, buried pins, intersecting pins, intersecting cells, 
intersecting nets, standard cell count, standard cell area, and area utilization. Figure~\ref{fig:pd_flow} shows step-by-step process of the proposed physical design flow.

\section{Experimental Results and Discussion}
\label{sec:analysis}

\subsection{Experimental Setup}
For the experiments and data collection, this study leveraged open-source, community-developed hardware from OpenCore~\cite{OpenCore:2019}. Those implementations feature various crypto cores, including MD5 pipeline, AES128, AES192, and AES256. Additionally, it incorporated an arithmetic core (FPGA median), processors such as OpenMSP and RISC16f84, a floating-point co-processor, and a video controller, namely JPEGEncoder. Logic synthesis and gate-level netlist extraction were performed using Synopsys Design Compiler (DC). The synthesis parameters include target clock frequencies (100 MHz to 2.5 GHz), aspect ratios (0.7 to 0.95), and core utilization (70\% to 95\%). Furthermore, the synthesis process employed Synopsys 28-nm CMOS technology library. After the logic synthesis, it uses Synopsys IC Compiler II for place \& route. The whole physical design flow was automated using tool command language (TCL) scripting. The research utilized about 60k data and divided the dataset for the proposed unsupervised model training, validation, and testing. The proposed training methodology uses 70\% DRC violation-free data to train and uses 15\% for
validation and 15\% for testing. This research only used 30\% DRC violated data for validation and the rest of the data for testing.
\begin{figure*}[t!]
	\begin{center}
		\vspace{-0.14cm}
		\includegraphics[width = 1.0\textwidth]{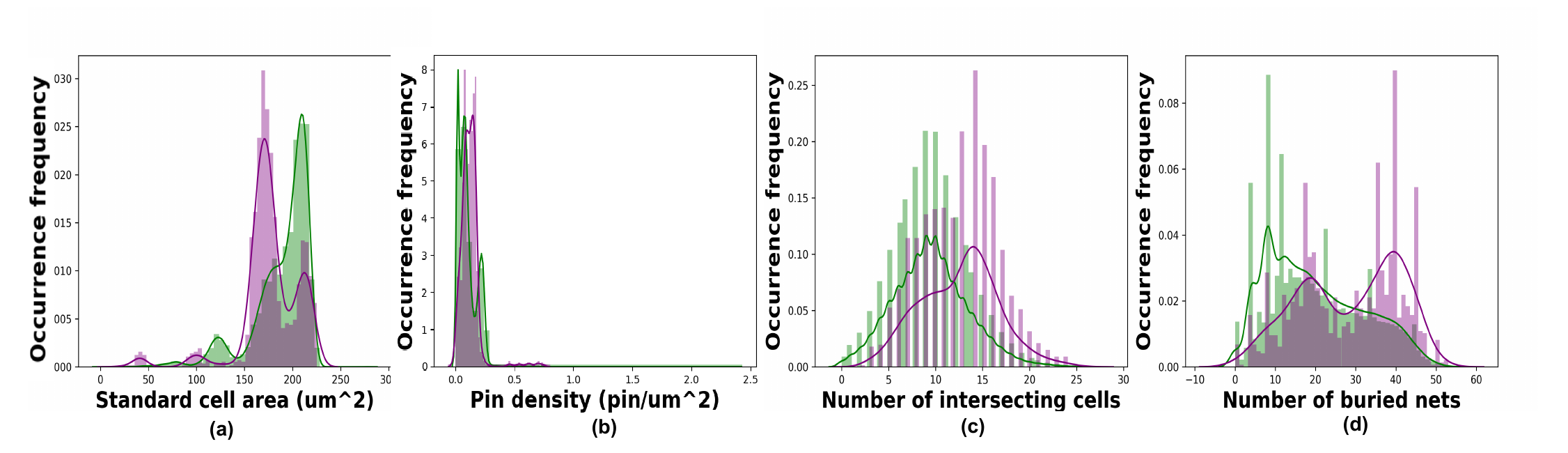}
		\vspace{-0.8cm}
		\caption {Gaussian distribution of DRCs considering (a) standard cell area, (b) pin density, (c) intersecting nets, and (d) buried nets.
			\label{fig:features_drits}}
		\vspace{-0.54cm}
	\end{center}
\end{figure*}

\subsection{Performance Validation and Comparative Analysis}
In order to validate the proposed algorithm, we fit each feature into a Gaussian distribution. Figure~\ref{fig:features_drits}(a),  Figure~\ref{fig:features_drits}(b), Figure~\ref{fig:features_drits}(c), and Figure~\ref{fig:features_drits}(d) show the Gaussian distribution considering standard cell area, pin density, intersecting nets, and buried nets, respectively. 
The green curve represents the DRC violation-free dataset, while the purple curve represents the data with DRC violation. 

The novelty of the proposed methodology is that we train using only DRC violation-free data. This research validated the training model with standard classification metrics like precision, recall, accuracy, and F1 scores. Table~\ref{tab:comparison} compares the proposed methodology with the state-of-the-art (SOTA) DRC violation prediction models considering validation and testing data. 
The proposed approach successfully detected all the DRC-violated data. The validation and test data showed a recall rate of 100\% in both cases and a precision of 99.43\% and 99.28\%, respectively. In addition, the proposed methodology has a 99.64\% F1 score, while the state-of-the-art machine learning approaches SVM classifier with RBF kernel~\cite{Chan_SVM:2016} has a 97.0\% and RF-based ensemble model has a 93.0\% F1 scores on the test data.

\begin{figure}[h!]
\centerline{\includegraphics[width = 0.5\textwidth]{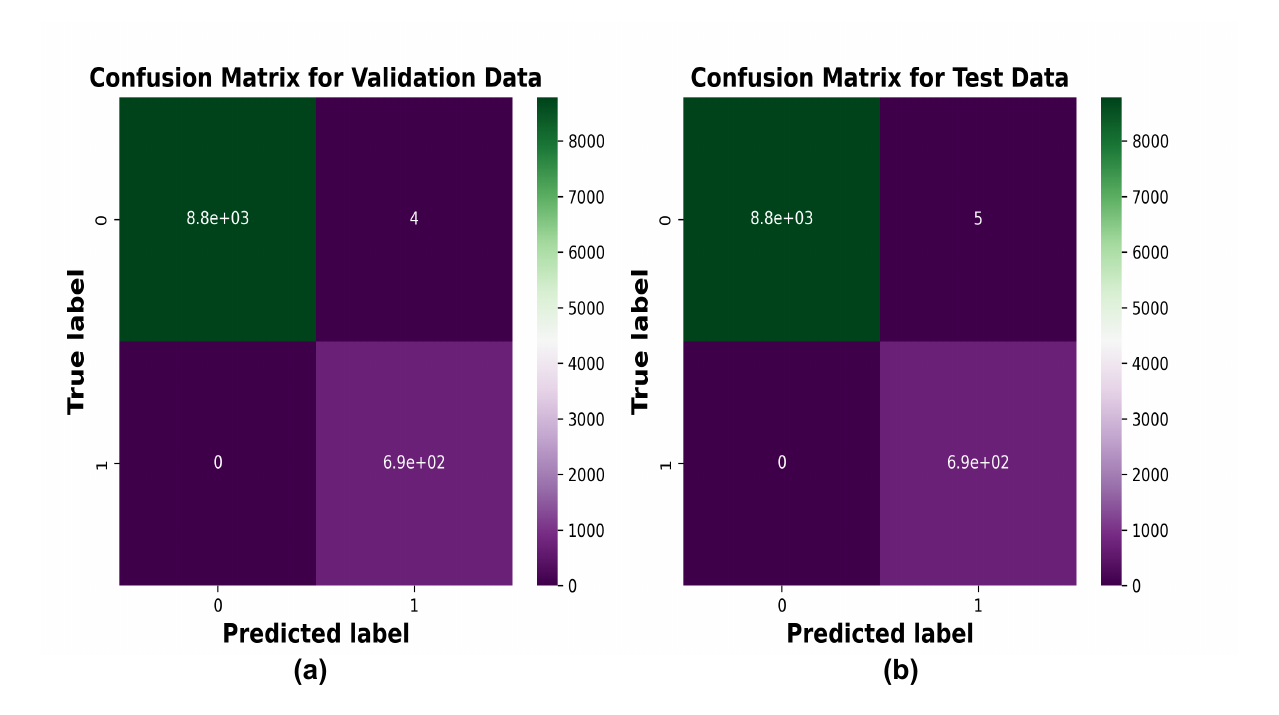}}
\vspace{-0.27cm}
\caption {For validation and test data, the proposed method has 99.96\% and 99.95\% accuracy compared to an NN-based recent model~\cite{pgnn_park:24}, which has a test accuracy of 98.74\%.}
\label{fig:confusion_mat}
\vspace{-0.05cm}
\end{figure}

This research compares the performance of the proposed model by comparing its predicted values with the actual values considering both validation and test data, as shown in Figure~\ref{fig:confusion_mat}. The proposed method has a zero false negative, implying that the model does not miss DRC violation cases. The proposed model has 99.96\% validation and 99.95\% test accuracy compared to recently reported PostGR CNN~\cite{cnn_hung:23} has 99.31\% and PA-GNN~\cite{pgnn_park:24} has 98.74\% test accuracies. 

\begin{table}[h]
    \centering
    \caption{Comparison of Different Methods for DRC Violation Prediction}
    \renewcommand{\arraystretch}{2.0} 
    \resizebox{\columnwidth}{!}{%
    \fontsize{14pt}{16pt}\selectfont 
    \begin{tabular}{|l|c|c|c|c|c|c|c|c|}
        \hline
        \multirow{2}{*}{\textbf{Methods}} & \multicolumn{2}{c|}{\textbf{Accuracy (\%)}} & \multicolumn{2}{c|}{\textbf{Precision (\%)}} & \multicolumn{2}{c|}{\textbf{Recall (\%)}} & \multicolumn{2}{c|}{\textbf{F1-score (\%)}} \\ 
        \cline{2-9}
        & \textbf{Val} & \textbf{Test} & \textbf{Val} & \textbf{Test} & \textbf{Val} & \textbf{Test} & \textbf{Val} & \textbf{Test} \\ 
        \hline
        SVM model~\cite{Chan_SVM:2016} & 99.81 & 85.44 & 98.2  & 98  & 98.3  & 98.0  & 97.5  & 97.0  \\ 
        \hline
        RF model~\cite{Islam_DAC:2019} & 95.88 & 85.88 & 99.2  & 99.0  & 88.9  & 88.0  & 93.8  & 93.0  \\ 
        \hline
        PostGR CNN model~\cite{cnn_hung:23}  & - & 99.31 & - & 80.57 & - & 77.22 & - & 79.12 \\ 
        \hline
        PreGR CNN model~\cite{cnn_jhen:23}  & - & - & - & 87.53 & - & 75.75 & - & 81.21 \\ 
        \hline
        PA-GNN model~\cite{pgnn_park:24}  & - & 98.74 & - & 72.2  & - & 72.12 & - & 72.16 \\ 
        \hline
        \textbf{Proposed Unsupervised} & \textbf{99.96} & \textbf{99.95} & \textbf{99.43} & \textbf{99.28} & \textbf{100} & \textbf{100} & \textbf{99.71} & \textbf{99.64} \\ 
        \hline
    \end{tabular}%
    }
    \label{tab:comparison}
\end{table}

{\bf Training Time:} Compared to the ML-based SVM model~\cite{Chan_SVM:2016}, the proposed unsupervised model has $26.3\times$ lower training time. Compared to other NN-based models, the proposed approach has $915\times$, $3420\times$, and $6003\times$ lower training times compared to PreGR CNN model~\cite{cnn_jhen:23}, PA-GNN model~\cite{pgnn_park:24}, and PostGR CNN model~\cite{cnn_hung:23}, respectively.

\section{Conclusion}
\label{sec:conclusion}

Accelerating the IC physical design flow is critical for the advancement of the microprocessor industry due to its potential to enhance performance, power efficiency, manufacturability, and competitiveness while reducing design cycles, costs, and complexity.
This research presented the first-ever unsupervised DRC violation prediction methodology. The proposed method has 99.95\% prediction test accuracy while the SOTA SVM, RF, and PA-GNN models have 85.44\%, 85.88\%, and 98.74\% accuracy, respectively.

Better yet, the proposed approach has 100\% Recall rate compared to SOTA RF, PostGR CNN, PreGR CNN, and PA-GNN, which have 88\%, 77.22\%, 75.75\%, and 72.12\% Recall rate, respectively.

Furthermore, the proposed methodology has about $26.3\times$ and up to $6003\times$ lower training times compared to SVM and PA-GNN models, respectively.

\bibliographystyle{IEEEtran}
\bibliography{main}

\end{document}